\documentclass[11pt,twoside]{article}
\usepackage{newpasp}
\def\edcomment#1{\iffalse\marginpar{\raggedright\sl#1\/}\else\relax\fi}
\reversemarginpar

\begin{document}
\title{AGN and Cooling Flows}
\author{James Binney}
\affil{ Oxford University\\
Theoretical Physics, 1 Keble Road, OX1 3NP}

\begin{abstract}
For two decades the steady-state cooling-flow model has dominated the
literature of cluster and elliptical-galaxy X-ray sources. For ten years
this model has been in severe difficulty from a theoretical point of view,
and it is now coming under increasing pressure observationally. 
A small number of enthusiasts have argued for a radically different
interpretation of the data, but had little impact on prevailing opinion
because the unsteady heating picture that they advocate is extremely hard to
work out in detail. Here I explain why it is  difficult to extract robust
observational predictions from the heating picture. Major problems include
the variability of the sources, the different ways in which a bi-polar flow can
impact on X-ray emission, the weakness of synchrotron
emission from sub-relativistic flows, and the sensitivity of synchrotron
emission to a magnetic field that is probably highly localized.

\end{abstract}

\section{Introduction}

Andy Fabian's talk was packed with images from the new generation of X-ray
telescopes. Images and spectra from Chandra and XMM will enliven the debate
between the proponents of the steady-state cooling-flow paradigm and those
who argue that unsteady heating by AGN is important. I want to mention some
things that need to be borne in mind if the debate is to be
fruitful. First let us recall the essentials of the cooling-flow model.

\section{The Cooling-Flow Paradigm}

The cooling-flow model was first outlined by Cowie \& Binney (1977) and was
elaborated by Fabian \& Nulsen (1977), Nulsen (1986) and others. Extensive
reviews can be found in Sarazin (1988) and Fabian (1994).

The X-ray emitting gas is considered to be in a quasi-steady state. It is extremely
inhomogeneous in that at any radius there is hot diffuse gas and cooler,
denser gas in approximate pressure equilibrium. The coolest gas rapidly
cools to very low temperatures and disappears from the X-ray sky. Its place
is taken by initially warmer gas, that cools to form cool gas. The place of
the warmer gas is taken by slightly warmer gas still, which cools to form
just warm gas, and so on right up to the hottest gas. Overall we have a
steady inward flow of material of different densities, which gains internal
energy as the gravitational field compresses it. Since in any comoving
volume of the matter there is a steady trickle of material `dropping out'
at extremely low temperatures and high densities, the density of the X-ray
observable gas does not rise as the gas moves inwards as rapidly as it would
if the gas were homogeneous.

By adjusting the radial profile of the mass dropout, or, equivalently, the
rate $\dot M(r)$ at which X-ray observable gas flows in across a sphere of
radius $r$, it is possible to obtain a perfect match to the
azimuthally-averaged X-ray surface-brightness profile. One finds that $\dot
M\propto r$ approximately. Once the brightness profile has been fitted to
the data and a metallicity chosen,  the X-ray spectrum at each radius can be
predicted. 

\section{Problems of the Cooling-Flow Model}

Prior to 1987 it was thought that Field's (1965) thermal instability would
naturally give rise to the multiphase medium that permits mass dropout.
Malagoli, Rosner \& Bodo (1987) showed that Field's analysis was entirely
misleading in a case such as this, in which the cooling medium has a
clear gradient in specific entropy and is in approximate hydrostatic
equilibrium in a gravitational field. In this case a cool, over-dense parcel
of gas (in which the specific entropy will be abnormally low) simply sinks
until it reaches the radius at which the bulk of the gas shares its specific
entropy.  Subsequently, the parcel executes vertical oscillations, that will
in practice be strongly damped by turbulence.

While some of us abandoned the cooling-flow model on reading Malagoli et al,
others went in with rubber bands to fix the model up. The rubber bands were
called magnetic field lines, and they pinned over-dense parcels in place.
Clearly, these field lines have to connect cooler gas to the hot embedding
matrix, and be fairly straight because they are under significant tension.
Consequently, heat will be efficiently conducted down them. Fields of the
expected strength can only pin clouds smaller in radius than $\sim20\,$pc,
and these will be evaporated away within a Hubble time unless the thermal
conductivity {\em along\/} field lines is several orders of magnitude
smaller than expected.

For some years the spectra predicted by the cooling-flow model have
conflicted with observation in that the data show fewer photons at low
energies, $\sim1\,$keV, than are predicted. This has been attributed to
`excess absorption' by cool gas, probably the very same gas that has dropped
out of the flow. The new generation of X-ray telescopes have shown that any
absorption is caused by gas at $\sim10^6\,$K, no cooler gas being
detectable (e.g., B\"ohringer et al., 2000). Searches at radio frequencies for emission by very cold gas have
repeatedly failed to detect significant reservoirs of gas, and very strong
limits have been set on  populations of young stars, from the highest masses
down to significantly below one solar mass (Prestwich et al., 1997).

Before the work of Malagoli et al., mass dropout seemed plausible in that the
inhomogeneities upon which it depended would be generated by thermal
instability from tiny seed inhomogeneities. Once one had read Malagoli et al,
it was apparent that one not only had to posit the existence of strongly
non-linear inhomogeneities in the initial conditions, but one also had to
take steps to prevent their prompt erasure. In these circumstances, it
seemed to some of us advisable to stand back and recall that cooling is
fastest where the density is highest, and this is known to be at the centre
of the cooling flow. Surely cooling to very low temperatures will occur first
at the centre? And will such cooling not feed the central black hole and
provoke it to an outburst? Might that outburst not restructure the X-ray
emitting gas, and thus regulate the cooling process?

\section{Heating by AGN}

This proposal that cooling flows are in a dynamical equilibrium with a
central AGN has the attraction of a fresh start. It sweeps away cobwebs in
the form of magnetically pinned over-densities and strange absorbing
screens. But one has to admit that at this stage it is little more than a
start: the detail is complex and formidably difficult to work out. Of course
we must seek critical confrontation with the data wherever possible, but we
must avoid damning the overall picture if we find that the data do not
support a naive expectation of what the model might predict.

Working out the observational consequences of AGN feedback on cooling flows
is hard because 

\begin{itemize}

\item the system is constantly far from a steady state;

\item the geometry is inherently complex;

\item to predict the  radio data one needs to know the distribution of
both the ultra-relativistic electrons and the magnetic field.

\end{itemize}

The cooling-flow model is very much simpler and has been rather completely
worked out. Hence the collision between the cooling-flow model and the
heating picture is asymmetrical: on the one side we have a model with
well defined predictions, and on the other a general picture from which it
is very hard to wring firm predictions. We rightly have a preference for the
simple over the complex, for the definite over the vague. Consequently, the
majority of people working in the field have clung tenaciously to the
cooling-flow model in the face of ever-increasing difficulties. Never the
less, I submit that the heating model enjoys a decisive advantage in another
important criterion by which we judge competing theories: a priori
plausibility in the light of our knowledge of the whole of physics and
astronomy. We must make a determined effort to work the picture up into a
well-explored model and judge it sternly only when we have extracted secure
predictions from it.

\subsection{Time-dependence}

The fundamental difference between the heating picture and the cooling-flow
model is that in the former partial derivatives with respect to time will
always be important. One expects an  individual source to cycle through a
series of stages. In the first stage the system cools towards a catastrophe.
In the second, the AGN is rejuvenated and blasts its immediate surroundings.
In the third turbulence dies down, and the entropy stratification of the gas
is restored before the first phase of the cycle resumes.

At each stage of the cycle one expects the source to look quite different.
The X-ray gas is expected to respond more slowly than the
synchrotron-emitting plasma, which will respond less quickly than any
central continuum source. In principle it should be possible to make
predictions about how many sources should be seen in each stage, but we
have neither such predictions nor the demographic knowledge that would be
required to test them. Proponents of the cooling-flow model are accustomed to
imagining the objects to be in steady states and one has to watch for them
implicitly carrying this assumption over to the heating picture.

Some important stages, such as that in which material cools to below
$1\,$keV, may be very short-lived and rarely observed. The phase of violent
energy output by the central black hole is likely to be subdivided into
extremely short-lived episodes, since the relevant dynamical time is a
matter of months only.

\subsection{Complex geometry}

Even if both the gravitational potential and the central heat source were
spherical, the resulting X-ray source would not be because heating from
below is inherently unstable: hot plasma will rise in some directions, while
cool plasma falls in others. Much of the energy released at the centre is
likely to emerge as ordered kinetic energy in a jet or other strongly
collimated outflow. One cannot credibly model the system in spherical
symmetry (though I have tried! -- see Binney \& Tabor, 1995). I am
skeptical that much useful can be achieved even in cylindrical symmetry (but
see D'Ercole \& Ciotti, 1998).
Fortunately, fully three-dimensional simulations are now just about
feasible, and we can hope for fairly realistic models to become available.
This will remain a computationally challenging problem for some time,
however, because the range of spatial scales and densities involved is
extreme -- from the parsec scale of the energy source to the megaparsec
scale of the X-ray halo; from a proton density in excess of
$10^{-8}\,$m$^{-3}$ near the centre of the cooling flow to densities many
orders of magnitude lower in the  plasma (e$_\pm$?) of a relativistic jet; from
inflow speeds of $\sim10\,$km$\,\hbox{s}^{-1}$ in the bulk of the cooling
flow to the speed of light in the jet. A further complication is the
possibility that the IGM becomes multiphase -- during the cooling phase this
will {\em not\/} occur, but gas heated from below {\em is\/} liable to
generate distinct phases, as has long been observed in the disk of the Milky Way.

One should beware simplistic assumptions about how a source's observable
properties will be affected by a jet. Displacement of thermal plasma by
relativistic plasma will suppress X-ray emission. On the other hand, X-ray
emission will be enhanced when thermal plasma is shocked or pushed up from
lower, cooler layers. Correspondingly, X-ray emission will be suppressed
when higher-entropy plasma sinks downwards to take up space left by the
upward movement of material. From the case of M87 we know too that the
synchrotron and inverse-compton processes can sometimes enhance X-ray
emission.

\subsection{Radio emission}

There is a great temptation to feel that one is seeing a true picture of the
relativistic plasma when one looks at a map of radio-continuum emission. In
fact, as Katherine Blundell (this volume) has emphasized, the synchrotron emissivity at a
given wavelength and for a given electron spectrum is a sensitive function
of the magnetic field strength. A simple argument suggests that the latter
is probably very non-uniform.

This argument starts from the observation that in a perfectly conducting
medium the magnetic field obeys the same equation as the vorticity (e.g.
Kulsrud et al., 1997).  Observations of rivers in spate and wind patterns
teach us that in high Reynolds number flows, vorticity is strongly
concentrated into sheets and vortex lines; in the language of non-linear
systems, vorticity is `intermittent'. We must expect magnetic fields in
turbulent conducting media to be similarly intermittent, and this
expectation is confirmed by recent observations of the Sun (Hagenaar et al.,
1999).  Hence, when we look at the delicate filamentary pattern of
synchrotron emission in an X-ray halo such as that of M87, we are probably
seeing merely the regions of enhanced $B^2$, while the relativistic
particles are more uniformly distributed. In particular, it is likely that
both X-ray emission and synchrotron emission derive from the same plasma;
the synchrotron emission comes from the high-energy tail of the particle
distribution, and is localized merely because $B^2$ is.

If it is true that ultra-relativistic electrons are space-filling out to the
outer boundary of the radio halo (which in the case of M87 lies at more than
half the cooling radius), substantial $P{\rm d}V$ work is likely to have
been done by the radio source on the surrounding thermal plasma.

 The radiating electrons are ultra-relativistic, and it is dangerous
to assume that these exotic particles dominate the total energy in
suprathermal particles; certainly in the solar neighbourhood, the cosmic-ray
energy is dominated by the least energetic particles. Suprathermal particles
with Lorentz factors $\gamma\sim1$ have extremely long lifetimes because
their synchrotron radiation is negligible and they are too fast to be
significantly Coulomb scattered. Consequently, they can persist and be
dynamically important long after synchrotron emission has become
unobservably faint.

Observations of jets suggest that the ultra-relativistic particles that
dominate radio maps are generated by jets with bulk Lorentz factors of a
few. These jets are unlikely to be produced by the accretion torus around
the AGN; they are probably produced by an exotic process, such as the
Blandford-Znajek (1977) effect, that taps the rotational energy of the black
hole. The fate of the  comparable or larger quantity of energy that is
released within the accretion disk has long been a puzzle. The ADAF model
(Narayan \& Yi, 1995)
posits that it is swallowed by the black hole. Blandford \& Begelman (1999)
have strongly criticised this model [which has recently encountered
difficulty matching observed spectral energy distributions (Di Matteo et
al., 2000)] and argue
that the energy is probably carried away by a wind off the accretion torus,
rather than swallowed by the black hole.  This conjecture chimes with
observations of accreting stars, from SS 433 to the sources of Herbig-Haro
objects, which imply that an accreting system is invariably associated with a
bipolar flow at velocities that range up to the largest Kepler velocity of
accreting material.

The implications for the heating picture of a significant fraction of the
accretion energy from an AGN emerging as a sub-relativistic bi-polar flow
would be considerable. The flow would probably be more steady than the
observationally much more conspicuous high-$\gamma$ jet that would
occasionally flicker to life at its centre. It would involve very much more
momentum than the high-$\gamma$ jet, and it would be hard to detect through
synchrotron radiation because in the sub-relativistic shocks that would
arrest it, very little energy would go into ultra-relativistic particles.
It is likely to generate structure in X-ray maps that is hard to trace in
radio maps because it is not associated with strong synchrotron emission.
Hence one should treat with caution claims that a weak correlation between
the radio and X-ray structures of an object imply that the AGN is not
heating the gas.

\end{document}